\documentclass[
reprint,
superscriptaddress,
bibnotes,
amsmath,amssymb,
aps,
prl,
longbibliography,
]{revtex4-2}

\usepackage{graphicx}% Include figure files
\usepackage{dcolumn}% Align table columns on decimal point
\usepackage{bm}% bold math
\usepackage{hyperref}% add hypertext capabilities
\usepackage{braket}
\usepackage{comment}
\usepackage{overpic} % in preamble
\usepackage{svg}

\usepackage{xcolor}

\newcommand{\qv}{\mathbf{q}}
\newcommand{\kv}{\mathbf{k}}
\newcommand{\av}[1]{\left\langle #1\right\rangle}

\begin{document}

%\preprint{APS/123-QED}

\title{Room-Temperature Metallic 2D Ferromagnetism Stabilized by Quantum Metric}% Force line breaks with \\

\author{Teng-Fei Ying}
\author{Jianqi Zhong}
\author{Songyuan Geng}
\author{Fangjie Chen}
\author{Haoxiang Li}
\author{Benjamin T. Zhou}
\email{tongz@hkust-gz.edu.cn}
\affiliation{Thrust of Advanced Materials \& Quantum Science and Technology Center,
The Hong Kong University of Science and Technology (Guangzhou), 511453, Guangzhou, China}

\date{\today}
 
\begin{abstract}

The recent discovery of long-sought dice-lattice bands in layered electride YCl~\cite{geng2026experimental} opens up unprecedented opportunities for studying correlated electrons in real dice lattice systems. While the weakly dispersive band at the Fermi level is partially filled and density-wave states are expected due to Fermi surface nesting based on conventional theories, the ARPES measurement nevertheless revealed a uniform flat band splitting of 0.5 eV, suggesting unusual ferromagnetism (FM) borne out of the \emph{metallic} 2D dice lattice. In this Letter, we propose that the uniform metallic FM order is stabilized by the prominent quantum metric near topological band crossing points at $K$ and $K'$, which reconstructs the magnetic instability in the dice metal and tips the density-wave instability toward a uniform FM order. Importantly, our analysis based on random phase approximation (RPA) and Hartree–Fock theory predicts a remarkably high Curie temperature above room temperature for the 2D FM order. These findings establish the first example of a room-temperature 2D metallic magnet stabilized by quantum metric, which not only exemplifies the fundamental significance of quantum geometry in solids but also bears far-reaching applications in energy-efficient high-density magnetic memories. 

\end{abstract}

\maketitle

\emph{Introduction.}---
Two-dimensional (2D) ferromagnetism sits at the heart of energy-efficient miniaturized magnetic memories and logic devices~\cite{burch2018magnetism,gong2019two}. The recent discovery of intrinsic ferromagnetism(FM) in van der Waals materials such as CrI$_3$~\cite{CrI3}, Cr$_2$Ge$_2$Te$_6$~\cite{Cr2Ge2Te6}, Fe$_3$GaTe$_2$~\cite{Fe3GaTe2,Fe3GaTe2_}, and Fe$_5$GeTe$_2$~\cite{Fe5GeTe2,Fe5GeTe2_,Fe5GeTe2__} has opened up possibilities for atomically thin magnets. In particular, Fe$_3$GaTe$_2$ and Fe$_5$GeTe$_2$ exhibit high Curie temperatures of 300 K, compatible with operations under ambient conditions, and the robust FM order arises from the enhanced density of states at the Fermi level through the well-established Stoner mechanism~\cite{tasaki1998nagaoka}. 

In parallel to efforts to raise the Curie temperature, another frontier in 2D ferromagnetism has revolved around flat-band ferromagnetism in moir\'{e} superlattice systems such as magic-angle twisted bilayer graphene and twisted transition-metal dichalcogenides~\cite{TBGFM,TBGFM1,TMDFM,TMDFM2, tschirhart2021imaging,cai2023signatures,park2023fractionally}. The strong Berry curvature and quantum geometric effects~\cite{yu2025quantum,gao2026quantum,liu2025quantum,torma2023essay}
%\textcolor{red}{[need the original references on quantum metric and quantum geometry]} 
embedded within the ultra-flat bands are found to contribute significantly to spin stiffness and magnetic ordering~\cite{twistqg,twistqgtmd,chen2026dipole,kitamura2026exact}.
The Curie temperatures of moir\'{e} magnets, on the other hand, remain stubbornly low, typically on the order of a few Kelvins~\cite{serlin2020intrinsic,TMDFM,TMDFM2}, largely due to the significantly diluted Coulomb interaction by the long moir\'{e} period. Consequently, quantum geometric effects in 2D ferromagnetism~\cite{oh2026magnetic,kitamura2026quantum} have been limited to date within the low‑temperature regime, far from practical room-temperature applications. This raises an important question: does quantum geometry play a fundamental role in 2D magnetism in the room-temperature regime?

\begin{figure}
\centering
\includegraphics[width=0.95\linewidth]{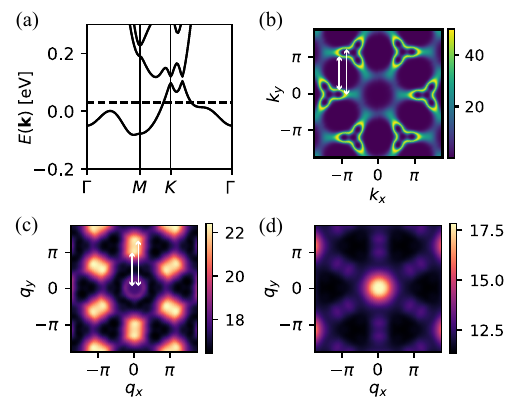}
\caption{(a) \textit{Ab initio}-based non-interacting band structure of YCl electride. The dashed line refers to the chemical potential at natural band filling. (b) Fermi surface of the realistic YCl model, obtained from the Fermi surface spectral function \(W_F(\mathbf{k})=\sum_n \left[-{\partial f(E_{n\mathbf{k}}-\mu)}/{\partial E}\right]\) at $T\simeq116$ K. (c) $\bm{q}$-resolved bare susceptibility $\chi^0_{s:\rm band}(\qv)$ with peak centered at finite ordering vector $\bm{Q} \simeq (0, \pi)$. White arrows in (b) and (c) indicate Fermi nesting vectors. (d) The full spin susceptibility $\chi^0_s(\qv)$ with quantum metric corrections resolved in $\bm{q}$-space. The peak is shifted to $\bm{q}=\bm{0}$.}
\label{fig:diceTBbandstructure_R}
\end{figure}

An ideal platform to address this question is a natural crystalline solid with an intrinsic topologically nontrivial flat band, where pronounced quantum metric and Berry curvatures can coexist with strong Coulomb interactions on the atomic scale. In a recent angle‑resolved photoemission spectroscopy (ARPES) measurement, the long-sought dice-lattice flat band was identified in the crystalline electride material yttrium monochloride (YCl)~\cite{geng2026experimental}, and the latest theoretical analysis has revealed the emergence of topological band crossings at the Brillouin zone corners $K$ and $K'$~\cite{zhong2025ycl}. A striking feature found in the ARPES spectrum is the presence of a giant energy splitting of approximately 0.5 eV between two sets of flat bands~\cite{geng2026experimental}, which points to a robust, long-range ferromagnetic (FM) order at $\qv = \bm{0}$. The compelling signature of uniform magnetic ordering, however, seems incompatible with the fermiology: at natural band filling, the weakly dispersive non-interacting band is partially filled in general (see representative band diagram in Fig.~\ref{fig:diceTBbandstructure_R}(c)) and the Fermi surface nesting condition with finite $|\qv|$ can be met under realistic conditions (Fig.~\ref{fig:diceTBbandstructure_R}(d)). This then raises the conceptual question: why does the system forgo the finite-$\qv$ density-wave instabilities favored by Fermi-surface nesting, but instead settle into a uniform $\qv = \bm{0}$ ferromagnetic ground state?

In this Letter, we show that the puzzle is resolved by the quantum metric embedded in the multi-orbital dice flat band in YCl. In particular, the prominent quantum metric near the topological band crossing points near $K$ and $K'$ dramatically reconstructs the magnetic instability of the system, shifting the peak of the magnetic susceptibility from finite-$\qv$ to $\qv = \bm{0}$, thereby stabilizing a uniform metallic ferromagnetic phase over competing density‑wave states. Importantly, our random phase approximation (RPA) analysis and Hartree–Fock theory predict a remarkably high Curie temperature above the room-temperature regime. These results establish the newly discovered YCl electride as the first material platform that exemplifies room-temperature metallic 2D ferromagnetism stabilized by quantum metric.

\emph{Spin susceptibility with quantum-metric corrections.}---
We start with a brief review of the bare Lindhard spin susceptibility $\chi^0(\qv)$, whose peak position determines the ordering vector $\bm{Q}$ of the magnetic phase when electron interactions are turned on within the Stoner-RPA formalism. The conventional Lindhard formula reads:
\begin{equation}\label{baresuscepband}
\setlength{\jot}{1pt} % default is usually 3pt
\begin{aligned}
\chi^0_{\mathrm{band}}(\qv,i\Omega_n)&=\sum_{n}\frac{1}{N_k}\sum_{\mathbf{k}}\frac{f(\epsilon_n(\kv+\qv))-f(\epsilon_n(\kv))}{\epsilon_n(\kv)-\epsilon_n(\kv+\qv)+i\Omega_n}\\
&=\sum_{n}\frac{1}{N_k}\sum_kF_{nn}(\kv,\qv).
\end{aligned}
\end{equation}
Here, \(f(\epsilon)\) is the Fermi distribution function and \(F_{nn}(\kv,\qv)\) denotes the conventional Lindhard factor. Note that Eq.~\eqref{baresuscepband} depends only on the band dispersion and ignores the detailed geometric structure embedded in the Bloch wave functions. While such an approximation holds reasonably well in a single-band, topologically trivial system, 
 form factors $\av{u_n(\kv+\qv)|u_m(\kv)}$ cannot be ignored in multi-orbital and particularly topologically nontrivial systems. The full expression for the bare susceptibility, which takes into account these form factors, is given by~\cite{kitamura2024spin}:
\begin{equation}\label{baresuscep}
\setlength{\jot}{1pt} % default is usually 3pt
\begin{aligned}
\chi^0(\qv,i\Omega_n)=&\sum_{nm}\frac{1}{N_k}\sum_{\mathbf{k}}\frac{f(\epsilon_n(\kv+\qv))-f(\epsilon_m(\kv))}{\epsilon_m(\kv)-\epsilon_n(\kv+\qv)+i\Omega_n}\\
&|\av{u_n(\kv+\qv)|u_m(\kv)}|^2\\
=&\sum_{nm}\frac{1}{N_k}\sum_{\mathbf{k}}F_{nm}(\kv,\qv)(1-D_{nm}(\kv,\qv)),
\end{aligned}
\end{equation}
where \(\epsilon_n(\kv)\) and \(\lvert u_n(\kv)\rangle\) are, respectively, the eigenvalue and eigenvector of the noninteracting Hamiltonian, and
\(
D_{nm}(\kv,\qv)
=1-\left|\av{u_n(\kv+\qv)|u_m(\kv)}\right|^2
\)
is the quantum distance between the corresponding Bloch states. This distance measures the dissimilarity between two Bloch states, thus encoding information of the underlying quantum geometry. 

The quantum-metric tensor is defined as the real part of the quantum-geometric tensor~\cite{yu2025quantum}: 
\begin{equation}
g^{\mu\nu}_{n}(\kv)=\operatorname{Re}\av{\partial_{k_\mu}u_n(\kv)\mid 1-P_n(\kv)\mid\partial_{k_\nu}u_n(\kv)},   
\end{equation} 
where \(P_n(\kv)=\ket{u_n(\kv)}\!\bra{u_n(\kv)}\) projects Bloch states onto the orthogonal subspace of \(\ket{u_n(\kv)}\). To explicitly reveal the role of quantum metric in $\chi^0(\qv,i\Omega_n)$, we consider without loss of generality the intra-band case with $n=m$, and the quantum distance under small-\(\qv\) expansion is expressed as $D_{n}(\kv,\qv)\simeq g_n^{\mu\nu}(\kv)q_\mu q_\nu$~\cite{yu2025quantum}. 
This clearly indicates that the quantum metric $g^{\mu\nu}_{n}(\kv)$ naturally enters the bare susceptibility and bears consequences in the magnetic instability. In particular, the curvature of static $\chi^0(\qv,\omega=0)$ around $\qv=\mathbf{0}$, defined as
$\chi_s^{0:\mu\nu} = \lim_{\qv \to \mathbf{0}} \partial_{q_\mu}\partial_{q_\nu}\chi_s(\qv)$
\footnote{The bare spin susceptibility is defined by
$
\chi^0_s(\qv,i\Omega_n)=\chi^0_{\uparrow\uparrow}(\qv,i\Omega_n)+\chi^0_{\downarrow\downarrow}(\qv,i\Omega_n)
-\chi^0_{\uparrow\downarrow}(\qv,i\Omega_n)-\chi^0_{\downarrow\uparrow}(\qv,i\Omega_n),
$
which could be further simplified as $\chi^0_s(\qv,i\Omega_n)=2\chi^0_{\sigma\sigma}(\qv,i\Omega_n)$ in the non-interacting regime.}, provides a useful criterion in inferring magnetic ordering behavior under interactions~\cite{kitamura2024spin,heinsdorf2025altermagnetic}: a positive trace of $\chi_s^{0:\mu\nu}$ creates a local minimum in $\chi_s(\qv)$ at $\bm{q} = \bm{0}$, which implies finite ordering vector $\bm{Q}\neq\bm{0}$; a negative trace of $\chi_s^{0:\mu\nu}$, in contrast, gives rise to a local maximum for $\chi_s(\qv)$ at $\bm{q} = \bm{0}$, and interactions tend to drive the system toward uniformly ordered FM phase with $\bm{Q}=\bm{0}$.

Notably, the quantum-metric always contributes to a negative trace of $\chi_s^{0:\mu\nu}$ and tends to stabilize a uniform FM order. This can be seen by noting that at small-$\qv$ the quantum-metric contribution to $\chi_s^{0:\mu\nu}$ can be written as: $\chi^{0:\mu\nu}_{\mathrm{geom;intra}}=\frac{2}{N_k}\sum_{n,\kv}f'(\epsilon_n(\kv))g^{\mu\nu}_n(\kv)$~\cite{kitamura2024spin,supp}. Since the diagonal part of the quantum-metric tensor is always non-negative: $g^{\mu\mu}_{n}(\kv)=\sum_{m\neq n}|\braket{\partial_{k_\mu}u_n(\kv)| u_m(\kv)}|^2\geq0$. Given $f'(\epsilon_n)\leq 0$, the trace of $\chi^{0:\mu\nu}_{\mathrm{geom;intra}}$ is necessarily non-positive.

\emph{Band structure, fermiology and magnetic susceptibility in YCl electride.}---
The non-interacting bands and Fermi surface of the monolayer YCl obtained from \textit{ab initio}-based multi-orbital tight-binding model~\cite{geng2026experimental, zhong2025ycl, supp} are presented in Fig.~\ref{fig:diceTBbandstructure_R}(a). The Fermi level at natural band filling is indicated by the dashed line, which intersects the weakly dispersive bands and results in two isolated, trigonally warped Fermi contours enclosing \(K\) and \(K'\) as shown in Fig.~\ref{fig:diceTBbandstructure_R}(b). 

Notably, due to the presence of nearly parallel Fermi contour segments connected by wave vectors \(\mathbf{Q}\) (white double arrows in Fig.~\ref{fig:diceTBbandstructure_R}(b)), the Fermi nesting condition is fulfilled, suggesting an ordering wave vector of $\bm{Q} \simeq (0, \pi)$ based on the conventional Lindhard formula in Eq.1 without quantum-metric corrections. Indeed, the spin susceptibility obtained using Eq.~\ref{baresuscepband} does exhibit six peak positions as shown in Fig.~\ref{fig:diceTBbandstructure_R}(c), which match perfectly with the Fermi nesting vectors in Fig.\ref{fig:diceTBbandstructure_R}(b). Thus, the conventional theory based on band dispersion only predicts a spin-density-wave order under interactions.

The magnetic instability changes dramatically under quantum-metric corrections. Using the full Lindhard formula in  Eq.~\ref{baresuscep} which takes into account the form factors, $\chi^0_s(\qv)$ is maximized at $\qv=\mathbf{0}$ (Fig.~\ref{fig:diceTBbandstructure_R}(d)) with a negative 
\(
\chi^{0:\mu\nu}_{s}<0
\). This clearly predicts a uniformly ordered FM state with $\bm{Q}=\bm{0}$. The contrasting results between Fig.~\ref{fig:diceTBbandstructure_R}(c) and (d) reveal the indispensable role of quantum-metric corrections in studying magnetic orders in the YCl electride system and dice-lattice in general.

\begin{figure}
    \centering
    \includegraphics[width=0.45\textwidth]{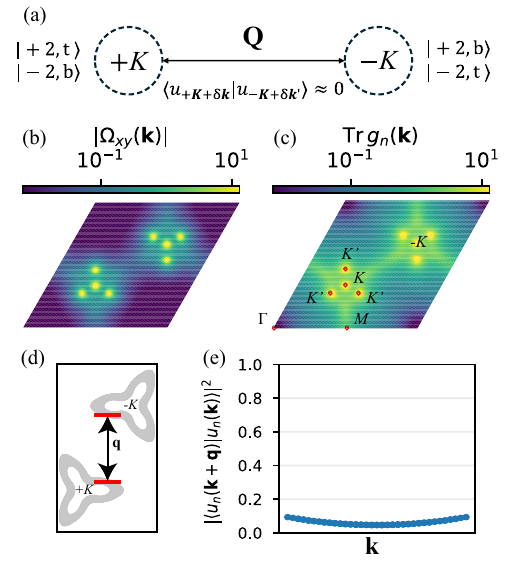}
    \caption{
    (a) Schematic of the layer-orbital-valley locking (LOVC) in YCl electride. Bloch states connected by the nesting vector $\bm{Q}$ are nearly orthogonal, thus suppressing the wave function overlap between Bloch states from opposite valleys aw well as the exchange coupling leading to density waves.
    (b) The Berry curvature distribution in the lowest two bands of YCl.
    (c) Trace of the quantum metric tensor $\sum_n[g_n^{xx}(\kv)+g_n^{yy}(\kv)]$ for the lowest two bands in YCl. 
    (d) Nested Fermi-surface segment at $\mu=0.03$ eV near \(\pm K\) connected by nesting vector \(\mathbf q\). 
    (e) Quantum-metric correction factor \(|\langle u_n(\mathbf k+\mathbf q)|u_n(\mathbf k)\rangle|^2\) with $\bm{q} \simeq \bm{Q}$ along the selected segment.
    }
    \label{fig:quantummetrictab-0.04}
\end{figure}

\emph{Origin of quantum-metric corrections: layer-orbital-valley coupling.}---
The physical origin behind the dramatic reconstruction in $\chi^{0:\mu\nu}_{s}$ due to quantum-metric corrections, as shown in Fig.~1(c)-(d), is rooted in the unique layer-orbital-valley coupling (LOVC) effect, as illustrated by Fig.~\ref{fig:quantummetrictab-0.04}(a) and discussed in detail in Refs.~\cite{zhong2025ycl}. The LOVC effect combined with atomic spin-orbit coupling in the $4d$-orbitals of yttrium atoms endows the multi-orbital dice flat band with a high Chern number of $\mathcal{C} = \pm 4$. In particular, the LOVC effect gives rise to Dirac band crossings near \(K\) and \(K'\) valleys, and the intrinsic SOC provides the Dirac mass terms, and giant Berry curvatures on the order of 10$^4$ {\AA}$^2$ emerge in the neighborhood of \(K\) and \(K'\) (Fig.~\ref{fig:quantummetrictab-0.04}(b)). Given the well-known trace condition \(\operatorname{Tr}g_n(\kv)\geq\left|\Omega_n^{xy}(\kv)\right|\)~\cite{yu2025quantum,roy2014band}, prominent quantum-metric effects also arise near the Berry curvature hot spots around \(K\) and \(K'\), thus leading to non-negligible quantum-metric corrections at $\bm{q} = 0$. The shift of the peak position from finite ordering vector $\bm{Q}$ to uniform FM order with $\bm{q} = \bm{0}$ is thus consistent with the general argument that the quantum-metric tends to enhance FM instability. This suppression of finite $\bm{Q}$ nesting due to LOVC is reminiscent of suppressed long-wave scattering due to destructive sublattice interference in some kagome systems~\cite{kitamura2026quantum}, but the LOVC differs from destructive interference as it is rooted in the orthogonality between different sublattice-orbital subspaces.

To draw a more concrete connection between LOVC and quantum-metric, we note that the LOVC separates Bloch states near \(K\) and \(K'\) into mutually orthogonal spaces. As the nesting vector $\bm{Q}$ (white arrows in Fig.1(b)) connect the two isolated Fermi pockets around \(K\) and  \(K'\), the two Bloch states involved in the form factors $|\av{u_n(\kv+\qv)|u_n(\kv)}|^2$ are nearly orthogonal for $\qv = \bm{Q}$ (Fig.~\ref{fig:quantummetrictab-0.04}(d)). This strongly suppresses the spin magnetic susceptibility at a finite nesting vector $\bm{Q}$ according to Eq.~\ref{baresuscep}. We confirm this observation numerically by plotting the form factors along the Fermi contour segment: the form factors are vanishingly small with $|\av{u_n(\kv+\qv)|u_n(\kv)}|^2 \leq 0.1$ for the nesting vector $\bm{Q}$ (Fig.~\ref{fig:quantummetrictab-0.04}(e)). From the quantum-geometric point of view, the LOVC maximizes the quantum distance (metric) between Bloch states involved in the Fermi nesting, thus suppressing the density-wave order.

\begin{figure}
    \centering
    \includegraphics[]{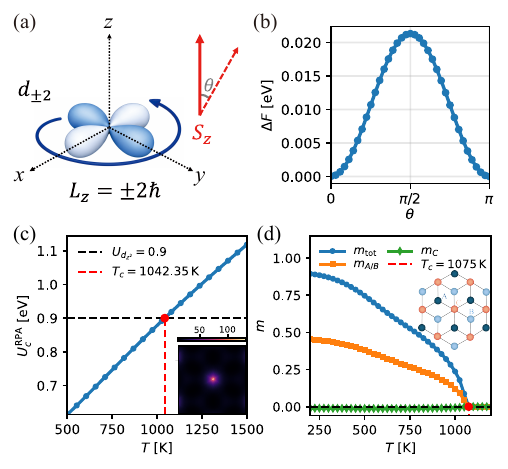}
    \caption{ 
    (a) Schematic of the $d_{\pm2}=d_{x^2-y^2}\pm i d_{xy}$ orbitals carrying an out-of-plane orbital angular momentum $L_z=\pm2\hbar$, which gives rise to the dominant atomic spin-orbit coupling in YCl~\cite{zhong2025ycl}. $\theta$ measures the angular deviation of spin-polarization from the $z$-axis.
    (b) Magnetic anisotropy energy per unit cell $\Delta F(\theta)$ at $T=300$~K. Due to the Ising-like spin-orbit coupling in YCl, the free energy is minimized for out-of-plane magnetization with $\theta=0$ and $\pi$, and maximized for in-plane magnetization with $\theta=\pi/2$.
    (c) The critical Hubbard interaction $U_c$ obtained under random phase approximation, where the horizontal line $U_{d_{z^2}}=0.9$ eV~\cite{zhong2025ycl} intersects $U_c(T)$ at a critical temperature as $T_c^{\rm RPA}\simeq1042$ K. Inner panel: the RPA susceptibility $\chi^{\rm RPA}_s(\qv)$ obtained with \textit{ab initio}-based Hubbard model of YCl with on-site Hubbard interaction $U=0.9$ eV at $T=1100$ K. (d) The total and sublattice-resolved magnetization $m_i=\frac{1}{2}(\av{n_{i\uparrow}}-\av{n_{i\downarrow}})$ versus temperature obtained under Hartree-Fock approximation at filling number $N=3.8$. The mean-field magnetic order parameter vanishes at a critical temperature $T_c^{\rm HF}\simeq1075$ K.}
    \label{fig:7band_chi_Tc}
\end{figure}

\emph{Circumvention of Mermin-Wagner theorem from magnetic anisotropy and room-temperature 2D ferromagnetism.}---
Having established the uniform FM order, we are now ready to discuss magnetic phase transitions driven by electron-electron interactions. As the flat band in YCl is borne out of a natural crystalline system without any superlattice engineering, the Coulomb energy scale is on the order of 1 eV~\cite{geng2026experimental,zhong2025ycl}, which far exceeds the typical effective Coulomb energy of order 10-100 meV in moir\'{e} bands. Therefore, we expect the Curie temperature of YCl to be orders of magnitude higher than typical values of 10K in moir\'{e} ferromagnets.

On the other hand, the well-celebrated Mermin-Wagner theorem~\cite{merminwagner}, which relies on the assumption of continuous symmetries, states that low-energy massless Goldstone modes prohibit the formation of long-range FM order in two space dimensions, and the true critical temperature for the magnetic phase transition in a 2D system with continuous SU(2) symmetry should be governed by the Berezinskii–Kosterlitz–Thouless (BKT) transition~\cite{BKT}. The realistic YCl electride material, however, exhibits strong intrinsic SOC of the form $H_{\mathrm{SOC}}\sim\lambda L_zS_z$ with $\lambda \simeq 20$ meV (shown schematically in Fig.~\ref{fig:7band_chi_Tc}(b)), which pins electron spins to the out-of-plane $z$-direction~\cite{zhong2025ycl}. This leads to large magnetic crystalline anisotropy energy on the order of 0.02 eV (Fig.~\ref{fig:7band_chi_Tc}(b)) and explicitly breaks the continuous SU(2) symmetry underlying the formation of massless Goldstone modes. The reduction from continuous SU(2) to discrete $Z_2$ symmetry thus circumvents the Mermin-Wagner theorem and allows the emergence of uniform FM order in 2D~\cite{CrI3,jenkins2022breaking}. This suggests that the magnetic phase transition in realistic YCl material belongs to the same universality class as those in the classic Ising model with only spin-up and spin-down configurations, which allows us to estimate $T_c$ through standard RPA formalism and Hartree-Fock mean-field calculations. 

We first determine the favored magnetic ordering vector and the transition temperature within the RPA framework. The interacting static spin susceptibility in the multiorbital basis reads:
\begin{equation}
 \chi^{\rm RPA}_s(\mathbf q, T)
=
\left[
1-\chi_s^0(\mathbf q, T)U
\right]^{-1}
\chi_s^0(\mathbf q, T),  
\end{equation}
where both the bare susceptibility $\chi_s^0(\mathbf q, T)$ and the Hubbard interaction 
$U$ are matrices in orbital space (see Supplementary Material~\cite{supp}). Note that the $T$-dependence in $\chi_s^0(\mathbf q, T)$ enters through the Fermi-Dirac functions. The magnetic instability occurs when the denominator vanishes, i.e., when the leading eigenvalue of $\chi_s^0(\mathbf q, T)U$ reaches unity. The wavevector $\qv$
at which this condition is first met determines the ordering vector, and the temperature at which it occurs defines $T_c$. For the realistic multi‑orbital model of YCl with the onsite Hubbard parameter $U_{d_z^2}$=0.9 eV derived from first‑principles estimates~\cite{zhong2025ycl}, we find that $\chi^{\rm RPA}_s(\mathbf q, T)$ remains sharply peaked at $\qv =0$ for all temperatures down to the critical temperature (in-set of Fig.~\ref{fig:7band_chi_Tc}(c)), and the Stoner condition yields a critical temperature of $T^{\rm{RPA}}_{c} \simeq 1042$K. 

To complement the RPA analysis with a more microscopic treatment of local magnetic moments on inequivalent sublattice sites, we perform self‑consistent Hartree–Fock (HF) calculations based on the same \textit{ab inito}-based Hubbard model of YCl and the same model parameters. The interacting Hamiltonian is solved self‑consistently by iterating the single-particle density matrix until convergence. The magnetic order parameter is defined as the net spin polarization per unit cell,
$m_i
=
\frac{1}{2}
\left(
n_{i\uparrow}-n_{i\downarrow}
\right)$,
where \(i\) labels the sites in the unit cell. 
In the mean-field approximation, the Curie temperature is defined by the point where the magnetic order parameter vanishes: $T_c^{\rm HF}\simeq 1075~\mathrm{K}$, in excellent agreement with the RPA result. As the self-consistent HF solution naturally incorporates the form factors in Eq.~\ref{baresuscep} in the Fock exchange functionals~\cite{zhou2022moire,hu2026ferromagnetism,bultinck2020ground,abouelkomsan2023quantum,soni2021multitude}, the HF results have naturally incorporated the quantum-metric corrections in the mean-field energy. 

Notably, while our RPA and HF framework neglects spatial and thermal spin fluctuations which may affect the exact value of Curie temperatures~\cite{torelli2019calculating}, the consistent high temperature scales obtained from the two independent approaches, \(T_c^{\rm RPA}\simeq 1042~\mathrm{K}\) and \(T_c^{\rm HF}\simeq 1075~\mathrm{K}\), indicate a robust uniform ferromagnetic order with transition temperatures a few times of the room temperature of 300 K. These fluctuations are expected to introduce only minor corrections to the Curie temperature, which should not affect the key conclusion of robust FM order under ambient conditions.

\emph{Conclusion and Outlooks.}---We resolved a fundamental conceptual contradiction between experiment and conventional theory for partially filled bands that neglects the quantum metric. The uniform FM order observed in ARPES at partial flat-band filling is impossible to explain within the standard Stoner or nesting‑driven picture, yet it emerges naturally once the quantum metric is properly taken into account. This result identifies the decisive, non-negligible role of the quantum metric in understanding quantum magnetism. A comparison with the recently investigated isostructural electride LaCl~\cite{geng2026lacl} further highlights this distinction - LaCl already exhibits a strong band-driven response near \(\mathbf q=0\)~\cite{supp}, whereas in YCl quantum geometry is decisive for ferromagnetism. 

These findings establish the newly discovered dice-lattice flat-band material YCl, to the best of our knowledge, as the first material realization of a quantum-metric-stabilized room-temperature 2D metallic ferromagnet. This discovery arrives at a moment when the demand for energy-efficient, high-density magnetic memory is rapidly growing, driven by the expansion of data‑centric technologies and artificial intelligence. Notably, the metallic character of the ferromagnetic ground state, combined with the nontrivial band topology, offers a particularly attractive route for all-electrical readout. As predicted in Refs.~\cite{sun2011nearly,tang2011high,zhong2025ycl}, the spin-orbit-coupling-induced gap at the Dirac points endows the weakly dispersive bands with a nonzero Chern number. The magnetization direction can thus be read out easily via the sign and magnitude of the Hall response. This built-in readout scheme is fully compatible with standard transport measurement protocols, substantially simplifying the architecture of magnetic memory cells. The robust perpendicular anisotropy, high spin polarization, topological properties, and ambient-condition accessibility make YCl electride a unique and promising platform for next‑generation miniaturized memory devices.\\

\emph{Acknowledgments.}---The authors thank Akito Daido, Jin-Xin Hu, Xuzhe Ying, Lu-Yao Zhong, Zhichun Ouyang, and K. T. Law for illuminating discussions. We acknowledge the support of NSFC-Young Scientists Fund (No.12504194, No. 12304093, No. 12447158), Guangdong Provincial Quantum Science Strategic Initiative (Grant No. GDZX2501004), Guangdong Provincial Talents Program 2025D03J0006, Guangzhou Basic and Applied Basic Research Scheme (No. 2024A04J4509), and Start-up Fund of HKUST(GZ) through Grant No. G0101000127 and No. G0101000263.

%\newpage

%\bibliographystyle{apsrev4-2} 

\bibliography{refer.bib}

\clearpage

\onecolumngrid
\setlength{\textheight}{24cm}
\setlength{\textwidth}{17cm}
\setlength{\parskip}{0pt}

\begin{center}

%{\Large \textbf{Supplemental Material for}}\\[0.5em]

{\large \textbf{Supplementary Material:\\ Room-Temperature Metallic 2D Ferromagnetism Stabilized by Quantum Metric}}\\[1em]
Teng-Fei Ying,
Jianqi Zhong,
Songyuan Geng,
Fangjie Chen, 
Haoxiang Li,
Benjamin T. Zhou$^*$\\[0.5em]

{\it
Thrust of Advanced Materials \& Quantum Science and Technology Center,
The Hong Kong University of Science and Technology (Guangzhou), 511453, Guangzhou, China
}\\[0.5em]

\end{center}

\vspace{1em}

\setcounter{section}{0}
\renewcommand{\thesection}{S\arabic{section}}

\setcounter{figure}{0}
\renewcommand{\thefigure}{S\arabic{figure}}

\setcounter{table}{0}
\renewcommand{\thetable}{S\arabic{table}}

\setcounter{equation}{0}
\renewcommand{\theequation}{S\arabic{equation}}

\setlength{\parindent}{2em}

\section{I. Realistic Seven-band Model for the YCl Electride }

In addition to the single-orbital dice lattice model, a multi-orbital model that captures the full symmetry and topology of the monolayer YCl~\cite{zhong2025intrinsic} was proposed. Here, we point out the quantum-geometric nature of the ferromagnetic phase in the interacting multi-orbital model with results for the magnetic susceptibility and quantum metric of the multi-orbital model.

\begin{figure}[h]
    \centering
    \includegraphics[width=0.35\linewidth]{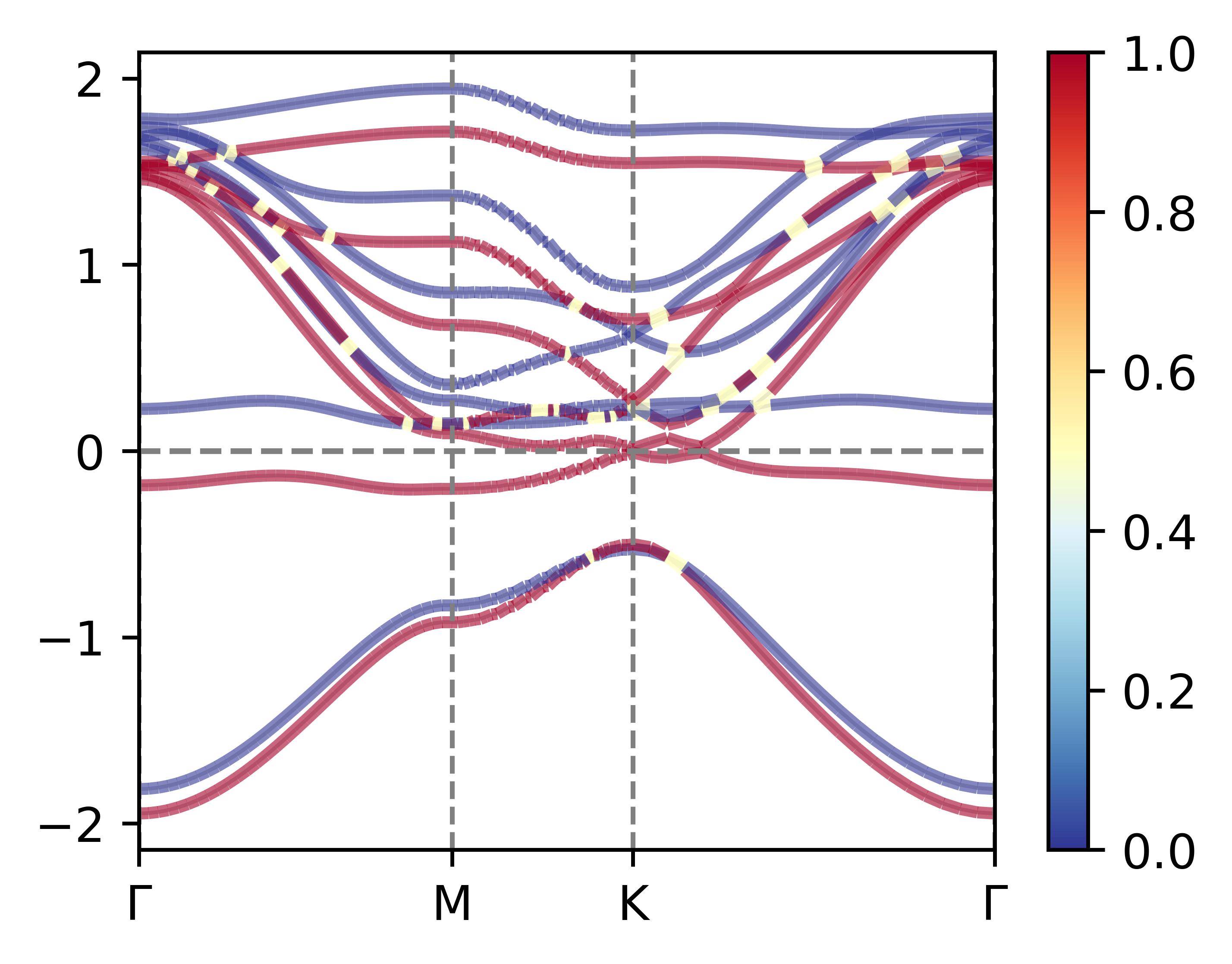}
    \includegraphics[width=0.35\linewidth]{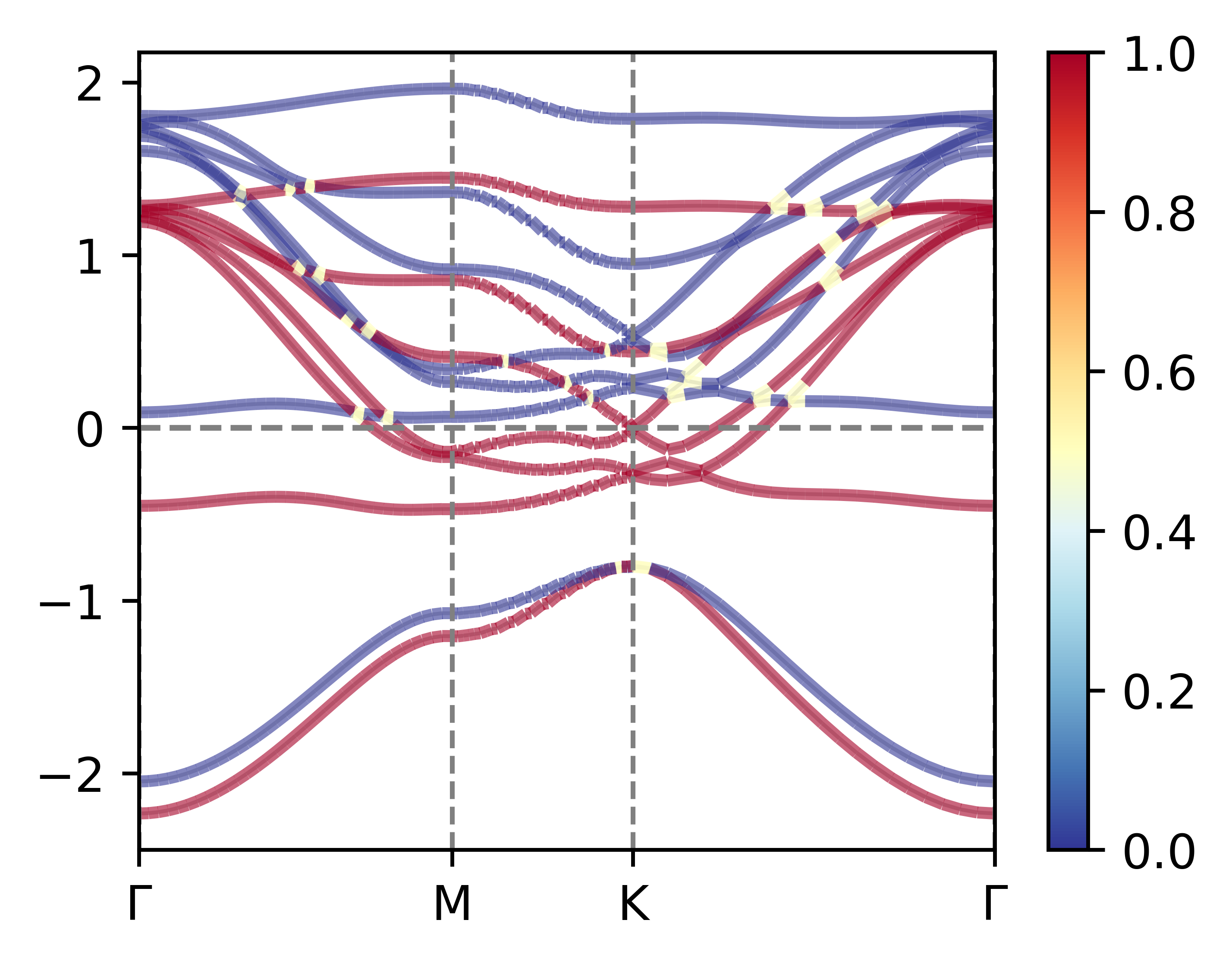}
    \caption{Spin-resolved Hartree-Fock band structure of the interacting realistic YCl model at $\beta=100$ eV$^{-1}$ and filling $N=3$ (left) and $N=3.8$ (right). The color scale indicates the spin-up weight of each eigenstate, with values from 0 to 1.}
    \label{fig:7bandHFN3}
\end{figure}

Different from the single-orbital model, where we assume that each sublattice only has the $s$ orbital, the A and B sites host the $d_{z^2}$ and $d_{x^2-y^2}\pm i d_{xy}$ orbitals, respectively, while the C site hosts the $s$ orbital in the multi-orbital model.
The corresponding magnetic quantum numbers are $m=0$ for the $s$ and $d_{z^2}$ orbitals and $m=\pm 2$ for the $d_{x^2-y^2}\pm i d_{xy}$ orbitals.
We therefore choose the Bloch basis
$
\left\{
\ket{\bm{k},\mathrm{A},0},
\ket{\bm{k},\mathrm{A},+2},
\ket{\bm{k},\mathrm{A},-2},
\ket{\bm{k},\mathrm{C},0},
\ket{\bm{k},\mathrm{B},0},
\ket{\bm{k},\mathrm{B},+2},
\ket{\bm{k},\mathrm{B},-2}
\right\},
$
from which the seven-band tight-binding Hamiltonian is constructed and subsequently extended to include spin. A full description of this multi-orbital model can be found in Ref.~\cite{zhong2025intrinsic}, where a ferromagnetic ground state is found with the interaction solved by the Hartree-Fock approximation. The Hartree-Fock band structure of the interacting model~\cite{zhong2025intrinsic} at $\beta=100$ eV$^{-1}$ (which corresponds to the temperature $T\simeq116$ K as $T=\frac{1}{k_B\beta}$, where $k_B$ is the Boltzmann constant) and filling \(N=3\) and \(N=3.8\) is shown in Fig.~\ref{fig:7bandHFN3}. The converged solutions of both cases correspond to the spin-polarized states along the \(z\) direction.

For the interacting part of the seven-band model, we consider local Coulomb interactions in the orbital basis. Since the seven orbitals have different spatial characters, the onsite interaction is taken to be orbital-dependent. In particular, the \(d_{z^2}\), \(d_{x^2-y^2}\pm i d_{xy}\), and \(s\)-orbital sectors are allowed to have different interaction strengths. In addition to the intra-orbital Hubbard interaction, we also include local inter-orbital density interactions between orbitals belonging to the same atomic site.

Therefore, the interacting part of the seven-band model is written as
\begin{align}
H_{\rm int}
=
\sum_i
\Bigg\{
\sum_{\alpha}
U_{\alpha}\,
n_{i\alpha\uparrow}
n_{i\alpha\downarrow}
\nonumber
+
\sum_{\langle\alpha,\beta\rangle}
\sum_{\sigma\sigma'}
\left(
U'_{\alpha\beta}
-
J\delta_{\sigma\sigma'}
\right)
n_{i\alpha\sigma}
n_{i\beta\sigma'}
\Bigg\}.
\label{eq:seven_band_interaction}
\end{align}
Here \(i\) labels the unit cell and \(\alpha,\beta\) label the seven local orbitals in the Bloch basis.

The notation \(\langle\alpha,\beta\rangle\) means that the inter-orbital density interaction is included only between different orbitals belonging to the same atomic site, namely within the \(A\)-site orbital group and within the \(B\)-site orbital group. Since the \(C\) site contains only one \(s\)-like orbital in the present basis, it has only the intra-orbital Hubbard interaction. With this convention, the opposite-spin inter-orbital interaction is \(U'_{\alpha\beta}\), while the same-spin interaction is reduced to \(U'_{\alpha\beta}-J\) due to the Hund coupling.
The interaction parameters used in the Hartree--Fock calculation are summarized in Table~\ref{tab:seven_band_interaction}.
\begin{table}[t]
\centering
\caption{Interaction parameters used in the seven-band Hartree--Fock calculation.}
\label{tab:seven_band_interaction}
\begin{tabular}{c c c}
\hline\hline
Orbital or interaction channel & Parameter & Value [eV] \\
\hline
\(\ket{\rm A/B,0}\), & \(U_{d_{z^2}}\) & \(0.9\) \\
\(\ket{\rm A/B,+2}\), \(\ket{\rm A/B,-2}\) & \(U_{d_{xy}}\) & \(2.5\) \\
\(\ket{\rm C,0}\) & \(U_s\) & \(0.3\) \\
\hline
Same-site inter-orbital, \(\sigma\neq\sigma'\) & \(U'\) & \(0.3\) \\
Same-site inter-orbital, \(\sigma=\sigma'\) & \(U'-J\) & \(0.2\) \\
Hund coupling & \(J\) & \(0.1\) \\
\hline\hline
\end{tabular}
\end{table}

\begin{figure}
    \centering
    \includegraphics[width=0.6\linewidth]{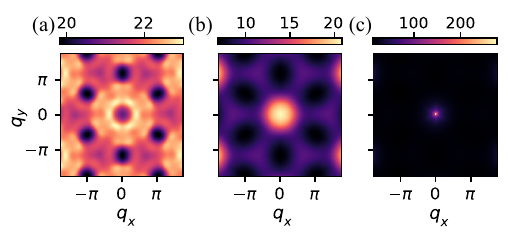}
    \caption{(a) The bare spin susceptibility of the realistic YCl model at the Fermi surface without the quantum geometry $\chi^0_{s:\mathrm{band}}(\qv)$ at $\beta=100$ eV$^{-1}$ and $\mu=0.00$ eV. White arrows in (a) and (b) indicate the possible Fermi nesting vectors. (c) The bare susceptibility from the same model as (b) but with the quantum-geometry effect $\chi^0_s(\qv)$. (d) The RPA susceptibility $\chi^{\rm RPA}_s(\qv)$ of the realistic YCl model at $\mu=0.00$ eV computed with onsite Hubbard interaction $U=0.9$ eV at $T=1100$ K.}
    \label{fig:7band_chi_Tc_sim}
\end{figure}

We further emphasize that, similar to the Hartree-Fock calculation, the quantum-metric-induced ferromagnetic fluctuation is not restricted to a single parameter point, but remains effective over a finite parameter range in the multi-orbital system. To demonstrate this, we present the bare susceptibility of the YCl model with and without the quantum-geometric effect, as well as the RPA susceptibility, at a different chemical potential, $\mu=0$ eV, in Fig.~\ref{fig:7band_chi_Tc_sim}, in comparison with Fig.~3 in the main text. The quantum-geometric contribution again reshapes the magnetic fluctuation and favors the ferromagnetic channel, showing that the mechanism is robust against changes in the chemical potential.

\section{II. Generalized Susceptibility}

Here, we show the derivation of the generalized susceptibility from 
\begin{equation}\label{baresuscep}
\begin{aligned}
\chi^0(\qv,i\Omega_n)=&\sum_{nm}\frac{1}{N_k}\sum_{\mathbf{k}}\frac{f(\epsilon_n(\kv+\qv))-f(\epsilon_m(\kv))}{\epsilon_m(\kv)-\epsilon_n(\kv+\qv)+i\Omega_n}
|\av{u_n(\kv+\qv)|u_m(\kv)}|^2\\
=&\sum_{nm}\frac{1}{N_k}\sum_{\mathbf{k}}F_{nm}(\kv,\qv)(1-D_{nm}(\kv,\qv)).
\end{aligned}
\end{equation}
Our goal is to compute the expression of $\chi^{0:\mu\nu}=\lim_{q\to0}\partial_{q_{\mu}}\partial_{q_\nu}\chi^0(\qv,i\Omega_n=0)$. To simplify the expression, we denote this as
\begin{equation}\label{GMSshort}
\lim_{q\to0}\partial_{\mu}\partial_{\nu}\chi(\qv)=\partial_\mu\partial_\nu\left\{\sum_{nm}\frac{1}{N_k}\sum_\kv F_{nm}(\kv,\qv)(1-D_{nm}(\kv,\qv))\right\}.
\end{equation}

After taking the partial derivatives, Eq.~\ref{GMSshort} becomes
\begin{equation}\label{partiald}
\begin{aligned}
\frac{1}{N_k}\sum_{nm,\kv}[&(\partial_\mu\partial_\nu F_{nm})(1-D_{nm}) 
- (\partial_\mu F_{nm})(\partial_\nu D_{nm})-(\partial_\nu F_{nm})(\partial_\mu D_{nm}) 
- F_{nm}(\partial_\mu\partial_\nu D_{nm})]|_{\qv\to 0} \\
=\frac{1}{N_k}\sum_{nm,\kv}&[(\partial_\mu\partial_\nu F_{nm})(1-D_{nm})-F_{nm}(\partial_\mu\partial_\nu D_{nm})]|_{\qv\to 0},
\end{aligned}
\end{equation}
with $\partial_\mu D_{nm}(\kv,\qv=0)=0$.

Because of the relation $\av{u_n(\kv+\qv)|u_m(\kv)}=\delta_{nm}$ at $\qv\to0$ so that $1-D_{nm}=\delta_{nm}$, the first term of Eq.~\ref{partiald} reduces to 
\begin{equation}\label{GMSband}
    \chi^{0:\mu\nu}_{\mathrm{band}}=\frac{1}{N_k}\sum_{n,\kv}\partial_\mu\partial_\nu F_{nn}(\kv,\qv\to0),
\end{equation}
which is only controlled by the band dispersion around the Fermi level instead of quantum geometry.

For the second term of Eq.~\ref{partiald}, we first consider the intraband part. The quantum distance can be related to the quantum metric as $D_{nn}(\kv,\qv)\sim g_n^{\mu\nu}(\kv)q_\mu q_\nu+\cdots$ when $n=m$~\cite{yu2025quantum}. Therefore, the partial derivative of the quantum distance can be written as $\partial_\mu\partial_\nu D_{nn}(\kv,\qv\to0)=2g_n^{\mu\nu}(\kv)$, which gives the intraband part of the geometry contribution:
\begin{equation}\label{GMSintra}
\begin{aligned}
\chi^{0:\mu\nu}_{\mathrm{geom;intra}}&=-\frac{2}{N_k}\sum_{n,\kv}F_{nn}(\kv,\qv\to0)g^{\mu\nu}_n(\kv)\\
&=\frac{2}{N_k}\sum_{n,\kv}f'(\epsilon_n(\kv))g^{\mu\nu}_n(\kv),
\end{aligned}
\end{equation}
using the relation $F_{nn}(\kv,\qv\to0)=-f'(\epsilon(\kv))$.

The interband part of the geometry contribution then is given by
\begin{equation}\label{geominter1}
\chi^{0:\mu\nu}_{\mathrm{geom;inter}}=-\frac{1}{N_k}\sum_{n\neq m,\kv}F_{nm}(\kv,\qv)\partial_\mu\partial_\nu D_{nm}(\kv,\qv)|_{\qv=0}.
\end{equation}
We write the quantum distance as $1-\av{u_n(\kv)|u_m(\kv+\qv)}\av{u_m(\kv)|u_n(\kv+\qv)}$, and take the partial derivative on this, Eq.~\ref{geominter1} becomes
\begin{equation}\label{geominter2}
\chi^{0:\mu\nu}_{\mathrm{geom;inter}}=\frac{2}{N_k}\sum_{n\neq m,\kv}F_{nm}(\kv,\qv)\Re[\av{\partial_\mu u_n|u_m}\av{u_m|\partial_\nu u_n}]|_{\qv=0}.
\end{equation}

Finally, taking Eqs.~\ref{GMSband}, \ref{GMSintra}, \ref{geominter2} together, we have the expression of the generalized susceptibility derived from Eq.~\ref{GMSshort}, as
\begin{equation}
\chi^{0:\mu\nu}=\chi^{0:\mu\nu}_{\mathrm{band}}+\chi^{0:\mu\nu}_{\mathrm{geom;intra}}+\chi^{0:\mu\nu}_{\mathrm{geom;inter}},
\end{equation}
which can be proved to be the same as Eq. S29 in Ref.~\cite{kitamura2024spin} by combining the Feynman-Hellmann theorem $\av{u_m|\partial_\mu u_n}=\frac{\av{u_m|\partial_\mu H_0|u_n}}{\epsilon_n-\epsilon_m}$.

\section{III. Magnetic Fluctuation}

Here, we followed the theory proposed by T. Kitamura et al to clarify the relation between the susceptibility and the ferromagnetism \cite{kitamura2024spin}.
Starting from the linear response theory, the spin susceptibility can be defined based on the Kubo formula by
\begin{equation}
\av{\hat{S^z}(\mathbf{r},t)}=S^z_0-\int d\mathbf{r'}\int^t_{-\infty}dt' \chi^s(\mathbf{r}-\mathbf{r'},t-t')h(\mathbf{r'},t').
\end{equation}
Here $\av{\hat{S^z}(\mathbf{r},t)}$ is the expectation value of the spin operator in the longitudinal ($z$) direction, $h(\mathbf{r},t)$ is the magnetic scalar potential, and $S^z_0$ is the original spin polarization.
After the Fourier transformation and the expansion of the scalar potential up to second order, we can obtain the equation for the spin polarization in the equilibrium state ($\omega\to0$), as 
\begin{equation}
\label{criterion}
\begin{aligned}
\delta\av{\hat{S^z}(\mathbf{r})}=-\chi_s(0)h(\mathbf{r})+\frac{1}{2}\lim_{\qv\rightarrow0}\partial_{q_\mu}\partial_{q_\nu}\chi_s(\qv)\partial_{r_\mu}\partial_{r_\nu}h(\mathbf{r}),\\
=-\chi_s(0)h(\mathbf{r})-\partial_{r_\mu}\left( \frac{1}{2}\lim_{\qv\rightarrow0}\partial_{q_\mu}\partial_{q_\nu}\chi_s(\qv)B_\nu(\mathbf{r})\right),
\end{aligned}
\end{equation}
where $\delta\av{\hat{S^z}(\mathbf{r})}=\av{\hat{S^z}(\mathbf{r})}-S^z_0$, and $\mu=x,y,z$.
From Eq.~\ref{criterion}, the criterion for the magnetic fluctuations is given by the structures of the bare spin susceptibility $\chi_s(\qv)$ around $\qv=0$, determined by the sign of $\chi_s^{0:\mu\nu}=\lim_{\qv\rightarrow0} \partial_{q_\mu}\partial_{q_\nu}\chi_s(\qv)$, the generalized magnetic susceptibility (GMS) extended from the generalized electric susceptibility (GES) in the charge susceptibility channel~\cite{heinsdorf2025altermagnetic}, is positive (antiferromagnetic fluctuation) or negative (ferromagnetic fluctuation). 

\section{IV. Susceptibility from Green's function}

The bare susceptibility can be calculated using the product of two single-particle Green’s functions in multi-orbital systems, as~\cite{tprf}:
\begin{equation}
\begin{aligned}
\chi^0_{ab}(\qv,i\Omega_n)&=\mathcal{F}_{\{\mathbf{r},\tau\}\rightarrow\{\qv,i\Omega_n\}}\left\{-G_{ba}(\mathbf{r},\tau)~G_{ab}(\mathbf{-r,-\tau})\right\},\\
&=-\frac{1}{N_k}\sum_{\kv,i\omega}G_{ba}(\kv,i\omega)~G_{ab}(\kv+\qv,i\omega+i\Omega),
\end{aligned}
\end{equation}
where $G_{ab}$ is the single-particle Green’s function in the orbital (sublattice) basis. To see the equivalence to the previous expression in Eq.~\ref{baresuscep}, one can write the Green's function in the band basis, 
\begin{equation}
G_{ab}(\mathbf{k},i\omega)
= \sum_{n} u_{a n}(\mathbf{k})\,u^{*}_{b n}(\mathbf{k})\,
G_{n}(\mathbf{k},i\omega),
\end{equation}
and insert into the bubble product, which gives
\begin{equation}
\begin{aligned}
\chi^{0}_{ab}(\mathbf{q},i\Omega)
= &\sum_{n,m}\frac{1}{N_{k}}\sum_{\mathbf{k}}
F_{nm}(\mathbf{k},\mathbf{q})
u_{a n}(\mathbf{k}+\mathbf{q})\,u^{*}_{a m}(\mathbf{k})\,
u^{*}_{b n}(\mathbf{k}+\mathbf{q})\,u_{b m}(\mathbf{k}).
\end{aligned}
\end{equation}
Noting the identity relation of the quantum state
\begin{equation}
\begin{aligned}
&\left|\braket{u_{n}(\mathbf{k}+\mathbf{q})|u_{m}(\mathbf{k})}\right|^{2}
=
\sum_{a,b}
u_{a n}(\mathbf{k}+\mathbf{q})\,u^{*}_{a m}(\mathbf{k})\,
u^{*}_{b n}(\mathbf{k}+\mathbf{q})\,u_{b m}(\mathbf{k}),
\end{aligned}
\end{equation}
which shows the total bare susceptibility obtained by summing the orbital (sublattice) components computed from Green's functions over $a,b$ is identical to the band-based Lindhard susceptibility in Eq.~\ref{baresuscep}:
\begin{equation}
\sum_{a,b}\chi^{0}_{ab}(\mathbf{q},i\Omega)
= \sum_{nm}\frac{1}{N_k}\sum_{\mathbf{k}}F_{nm}(\kv,\qv)(1-D_{nm}(\kv,\qv)).
\end{equation}

Therefore, one can analyse the susceptibilities with specific sublattice indices to isolate the fluctuations associated with particular ordering patterns. For example, consider ferrimagnetic configurations in which some sites align while others are oppositely polarized in the bipartite lattice.

\section{V. Connection to the Wannier-function Spread}

The physical meaning of the quantum-geometry-induced magnetism can be understood in terms of the Wannier-function spread in real space. 

The definition of the Wannier function in real space is
\begin{equation}
\ket{\mathbf{R}_n}=\frac{V}{(2\pi)^3}\int d\mathbf{k} e^{-i\mathbf{k\cdot R}}\ket{\psi_{n\mathbf{k}}},
\end{equation}
with the Bloch function $\psi_{n\mathbf{k}}(\mathbf{r})=e^{i\mathbf{k\cdot r}}u_{n\mathbf{k}}(\mathbf{r})$. Though there is an arbitrariness in the definition of Wannier functions, we can still calculate the spread (delocalization) of Wannier functions in real space by the momentum space overlap matrices as $\Omega=\Omega_I+\tilde{\Omega}$~\cite{marzari1997maximally}. Here, $\tilde{\Omega}=\Omega_{\rm OD}+\Omega_{\rm D}$ is the gauge-dependent part that is supposed to be minimized in practice, and $\Omega_I$ is the gauge-invariant part defined by
\begin{equation}
\Omega_I=\frac{1}{N_k}\sum_{\kv,\qv}\omega_\qv \sum_{nm}\left( \delta_{nm}-|M_{nm}(\kv,\qv)|^2\right),
\end{equation}
where $\omega_\qv$ is the numerical associated weight~\cite{marzari1997maximally}, and $M_{nm}(\kv,\qv)$ is defined by $M_{nm}(\kv,\qv)=\av{u_n(\kv+\qv)|u_m(\kv)}$. By definition, we note that the matrix $|M_{nm}(\kv,\qv)|^2$ is nothing but $1-D_{nm}(\kv,\qv)$. Therefore, we can rewrite the gauge-invariant part of the spread of Wannier functions by
\begin{equation}\label{spreadI}
\Omega_I=\frac{1}{N_k}\sum_{\kv,\qv}\omega_\qv \sum_{nm}\left( \delta_{nm}-(1-D_{nm}(\kv,\qv))\right),
\end{equation}
which is controlled by the same Bloch-state overlap loss as the quantum distance $D_{nm}(\kv,\qv)$. 

By comparing Eq.~\ref{spreadI} to Eq.~\ref{baresuscep}, we note that the same overlap-loss factor controlling the gauge-invariant Wannier spread also enters the bare susceptibility. Therefore, a more delocalized Wannier function, corresponding to a larger overlap mismatch in momentum space, generally produces a stronger quantum-geometric correction to the susceptibility.
Since the generalized magnetic susceptibility is determined by the curvature of the susceptibility at $\qv=0$, this geometric contribution can reshape the balance between ferromagnetic and antiferromagnetic fluctuations.

\section{VI. Hartree--Fock Approximation}

To verify whether the quantum-geometry-enhanced ferromagnetic fluctuation can develop into a stable magnetic ordered state, we study the dispersive dice-lattice and realistic YCl Hubbard models~\cite{zhong2025intrinsic} at finite temperature within the Hartree-Fock (HF) approximation. 

We consider the local Hubbard interaction in the orbital basis and allow the interaction strength to depend on the orbital character. This is physically motivated by the fact that the flat band near the Fermi level has dominant \(d_{z^2}\)-orbital weight, while the remaining orbitals mainly contribute through hybridization. The HF decoupling captures how the interaction acts most strongly on the orbitals forming the flat-band states. During the self-consistency loop, the spin-dependent onsite potential is updated from the local density matrix until the charge and spin densities converge.
The converged self-consistent density matrix determines the final magnetic order. From the converged solution, we extract the site- and spin-resolved occupations $\langle n_{i\sigma}\rangle$,
and define the local magnetic moment as
\begin{equation}
m_i=\frac{1}{2}\left(
\langle n_{i\uparrow}\rangle
-
\langle n_{i\downarrow}\rangle
\right).
\end{equation}
The sublattice-resolved moments on the A, B, and C sites are then used to identify the magnetic character of the HF solution. 
Therefore, the HF calculation gives a complementary confirmation that the quantum-geometry-enhanced ferromagnetic fluctuation can be promoted into a spin-polarized ordered phase in the interacting regime.

\section{VII. Random Phase Approximation}

In this section, we describe the random phase approximation (RPA) procedure used to analyze the magnetic instability of the multiorbital tight-binding model.

For simplicity, we only consider the local onsite Hubbard interaction
\begin{equation}
H_{\rm int}=U\sum_{i,\alpha} n_{i\alpha\uparrow}n_{i\alpha\downarrow},
\end{equation}
where \(\alpha\) labels the orbital or sublattice degree of freedom. Starting from the noninteracting Hamiltonian, we first compute the static bare susceptibility \(\boldsymbol{\chi}^0(\mathbf q)\) in the orbital basis. Here \(\boldsymbol{\chi}^0(\mathbf q)\) is a matrix whose elements $\chi_{\alpha\beta}$ describe the response between different orbital channels.

Within RPA, the spin susceptibility is obtained as
\begin{equation}
\boldsymbol{\chi}^{\mathrm{RPA}}_s(\mathbf q)
=
\left[
\mathbf 1-\boldsymbol{\chi}^0(\mathbf q)\mathbf U
\right]^{-1}
\boldsymbol{\chi}^0(\mathbf q),
\label{eq:rpa_matrix_simple}
\end{equation}
where \(\mathbf U=U\otimes\mathbf1\) for an orbital-independent onsite interaction. 
For the full susceptibility, we reduce the matrix susceptibility to a scalar quantity by contracting over the orbital indices,
\begin{equation}
\chi^{\mathrm{RPA}}_s(\mathbf q)
=
\sum_{\alpha,\beta}
\left[
\boldsymbol{\chi}^{\mathrm{RPA}}_s(\mathbf q)
\right]_{\alpha\beta},
\end{equation}
which corresponds to the response of the total spin density summed over all orbitals. 
Therefore, the magnetic instability occurs when the RPA denominator develops a zero eigenvalue. 
In practice, for each momentum \(\mathbf q\), we diagonalize the Hermitian part of \(\boldsymbol{\chi}^0(\mathbf q)\) and denote its largest eigenvalue by \(\lambda_{\max}(\mathbf q)\). 
The critical interaction at momentum \(\mathbf q\) is then
\begin{equation}
U_c(\mathbf q)=\frac{1}{\lambda_{\max}(\mathbf q)}.
\end{equation}
The leading instability is determined by the largest susceptibility eigenvalue in the Brillouin zone:
\begin{equation}
U_c=
\frac{1}{\max_{\mathbf q}\lambda_{\max}(\mathbf q)}.
\label{eq:Uc_simple}
\end{equation}
The corresponding ordering wave vector is the momentum \(\mathbf q^\ast\) where \(\lambda_{\max}(\mathbf q)\) is maximal. 
If \(\mathbf q^\ast=\Gamma\), the leading instability is ferromagnetic; if \(\mathbf q^\ast\) is finite, the leading instability corresponds to an antiferromagnetic or charge/spin-density-wave fluctuation.

\section{VIII. Magnetic susceptibility of LaCl}

\begin{figure}
    \centering
    \includegraphics[width=0.9\linewidth]{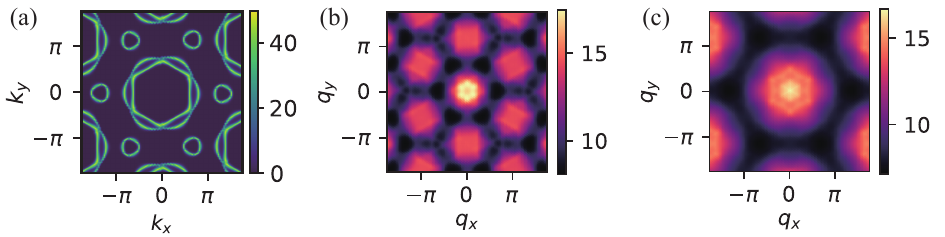}
    \caption{(a) Fermi surface of the LaCl model~\cite{geng2026lacl}, obtained from the Fermi surface spectral function \(W_F(\mathbf{k})=\sum_n \left[-{\partial f(E_{n\mathbf{k}}-\mu)}/{\partial E}\right]\) at $T\simeq116$ K and \(\mu=0\) eV. (b, c) The bare spin susceptibility of the realistic multi-orbital LaCl model at the chemical potential $\mu=0$ eV and $T\simeq116$ K. Panel (b) refers to the bare susceptibility with only the band-dispersion effect $\chi^0_{s:\rm band}(\qv)$, and panel (c) refers to the bare susceptibility with the quantum geometry $\chi^0_s(\qv)$.}
    \label{fig:lacl}
\end{figure}

LaCl is a recently identified electride that shares the same crystal structure as YCl but differs from it in many other respects~\cite{geng2026lacl}. Figure \ref{fig:lacl} shows that the band-only susceptibility \(\chi^{0}_{s,\mathrm{band}}(\mathbf q)\) of LaCl exhibits a broad maximum near \(\mathbf q=0\), indicating a predominant tendency toward long-wavelength magnetic order. Including the quantum-geometric contribution enhances and sharpens the response at \(\mathbf q=0\), thereby favoring uniform ferromagnetism (see Ref.~\cite{geng2026lacl} for the model we used to calculate for LaCl). This contrasts with YCl, where quantum geometry plays the decisive role in generating and selecting the ferromagnetic response. Thus, although both materials ultimately favor ferromagnetism, their underlying microscopic mechanisms are distinct, which is a difference that calculations identifying only the magnetic ground state cannot resolve.

\section{End Matter}

\begin{figure}
    \centering
    \includegraphics[width=0.98\linewidth]{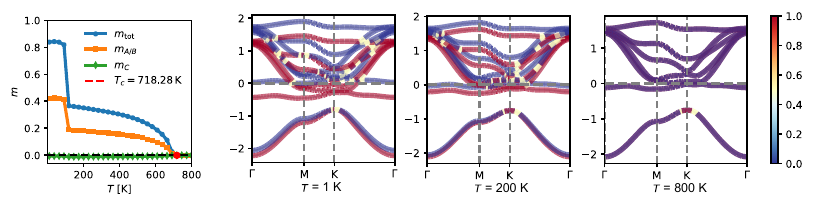}
    \caption{The magnetization $m_i=\frac{1}{2}(\av{n_{i\uparrow}}-\av{n_{i\downarrow}})$ computed from the Hartree-Fock approximation with smaller interaction ${\bf U}=0.885{\bf U_0}$ than our benchmark and the filling number $N=3.8$.}
    \label{fig:7bandHFN3.8}
\end{figure}

An experimental study attributed the kink near 100 K in the temperature-dependent magnetization of the two-dimensional YCl electride to a Curie transition~\cite{song2021van}. Our Hartree-Fock results suggest a different interpretation. Since mean-field Hartree-Fock generally overestimates the stability of ordered phases, we use reduced interaction parameters, \(\mathbf{U}=0.885\mathbf{U_0}\), where $\bf U_0$ is the parameter set given by Tab.~\ref{tab:seven_band_interaction}, to compare with the experimental magnetization curve, as shown in Fig.~\ref{fig:7bandHFN3.8}. The calculated magnetization exhibits a sharp step-like reduction near 100 K, resembling the experimental anomaly. However, the magnetization remains finite above this temperature, and the band structure, e.g., at \(T=200\) K, still shows clear spin-dependent reconstruction compared with the nonmagnetic state. Thus, the 100 K anomaly is more naturally interpreted as a spin-state transition or crossover within the ordered phase, rather than as the Curie temperature. In our calculation, the magnetic order is lost only at a much higher temperature, \(T_c \simeq 718\) K, where the magnetization vanishes, and the spin-resolved band reconstruction disappears.

\end{document}